\documentclass[prb,superscriptaddress,floatfix,twocolumn,showpacs,amsfonts,amsmath,amssymb]{revtex4}
\usepackage{graphicx} 
\usepackage{dcolumn} 
\usepackage{bm} 
\usepackage{color}

\begin{document}

\title{Magnetic focussing of electrons and holes in the presence of spin-orbit interactions}

\author{Samuel Bladwell}
\affiliation{School of Physics, University of New South Wales, Sydney 2052, Australia}
\author{Oleg P. Sushkov}
\affiliation{School of Physics, University of New South Wales, Sydney 2052, Australia}

\begin{abstract}

In this work we theoretically investigate transverse magnetic focussing in two dimensional electron and hole gasses with strong spin orbit interactions. We present a general result for spin orbit interaction with singular winding numbers in the adiabatic limit. We then present results for systems with two spin orbit interactions of different winding number appear, using the concrete and experimentally relevant case of  an applied in-plane magnetic field in hole systems with Rashba type interactions. We predict that the application of a large in-plane field is found to have a strong effect on the magnetic focussing spectrum.

\end{abstract}

\pacs{72.25.Dc, 71.70.Ej, 73.23.Ad  }
\maketitle

\section{Introduction}
\begin{figure}[t!]
{\includegraphics[width=0.3\textwidth]{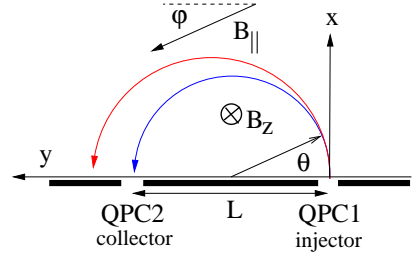}}
\caption{In magnetic focussing, electrons (or holes) are focussed with weak 
magnetic field from an injector quantum point contact (QPC) to a collector QPC,
at a distance $L$. }
\label{F1}
\end{figure}

The dynamics of charge carriers in spin-orbit coupled systems is a vital 
area of investigation for the extremely active field of spintronics. 
Controlling and manipulating the flow of electrons and holes serves as the 
foundation of an entire class of spintronic devices, most notably the 
Datta-Das spin transistor and its progeny \cite{Datta1990}. 
Experimental studies of spin-orbit systems, with an eye to these potential 
applications, are extensive. One particularly fruitful experimental technique 
is transverse magnetic focussing (TMF), which involves the mesoscopic 
transport of charge carriers from an injector quantum point contact (QPC) to 
a collector QPC, in a two dimension (2D) charge gas, through which the charge 
carrier can propagate ballistically on a scale of tens of microns.
The charge carriers are ``focused'' by out of plane magnetic field $B=B_z$
as illustrated in Fig. \ref{F1}. The two possible trajectories
shown in Fig. \ref{F1} correspond to two possible spin polarisation.
The figure illustrates that the problem under investigation is how the spin 
influences the semiclassical long-range orbital dynamics.
As an experimental technique, TMF has provided important insights into the 
properties of heterostructures and Quantum point contacts \cite{Vanhouten1989}.
More recently, TMF has been applied in Graphene \cite{Taychatanapat2013}. 
Another recent development is the use of two dimensional systems with large 
spin orbit (SO) interactions  in TMF, the goal to observe spin splitting in the 
magnetic focussing spectrum. Such a splitting has been observed in heavy hole 
systems using zinc blend heterostructure \cite{Rokhinson2004}. 
Simultaneously there has been extensive theoretical investigation of TMF
with Rashba SO interaction which is linear  in the particle momentum
using a variety of exact numerical \cite{Schliemann2008}, and strong and 
weak coupling semiclassical approximations \cite{Zulicke2007,Usaj2004}. 
In addition, polarised photocurrents have been experimentally detected using 
TMF \cite{Li2012}, and TMF has been used as an efficient method for detecting 
density differences in spin species in a QPC \cite{Reynoso2007, Rokhinson2006}. 

While extensive, theoretical investigation so far has been limited to the 
linear in the particle momentum SO interactions, like usual Rashba  interaction. 
Though often dominant in two dimensional electron systems, for heavy hole 
based systems the spin-orbit Rashba-like interaction is cubic in momenta. 
Particular crystal lattice orientations with respect to the 2D heterstructure
in zinc-blend  semiconductors result in cubic in momenta Dresselhaus 
SO interaction \cite{Winkler2003}. 
Application of an in-plane magnetic field to a heavy hole zinc-blend  
semiconductor heterostructure results in an effective SO interaction which is 
quadratic in the particle momentum ~\cite{Tommy}.
Furthermore, most 2D systems contain SO interactions of 
different orders in momentum simultaneously. Such a combination can suppress or enhance 
the spin splitting,
depending on the growth orientation and confinement shape. Such partial, or 
complete compensation of spin orbit interactions forms the basis of many 
proposed spintronic devices. 

In light of these considerations, in this work we pursue a theoretical 
investigation of TMF for a wide variety of SO interactions
which scales as some power of the particle momentum.
Moreover, we consider a case when SO interaction with different powers of momentum
act simultaneously.
We assume that the focusing magnetic field is weak and hence the evolution
of the particle wave function is adiabatic. 

The paper is structured as follows. In section \ref{sec1}
we present theory for SO interactions with a single winding number, and determining the peak position in
a magnetic focussing experiment. In section \ref{sec2} we extend this to the case of 
two SO interactions acting simultaneously, and determine the change to the magnetic focussing spectrum.
Finally, in section \ref{sec3} we briefly examine the question of injection via the QPC with 
a greatly simplified model, determining the angular distribution.

\section{Spin-orbit interactions with a given winding number} 
\label{sec1}

Semiconductor heterostructures allow for a great diversity of SO interactions. To render our 
results as more general, we consider the following spin orbit interaction for a 
two dimension electron or hole gas (2DEG),
\begin{eqnarray}
{\cal H}_{so} = i \frac{\gamma}{2} p_-^n \sigma_+ +h. c.
\label{spinorbit1}
\end{eqnarray}
Here $p$ is the particle momentum, $p_{\pm}=p_x\pm i p_y$, and
$\sigma_\pm = \sigma_x \pm i \sigma_y$, are the Pauli matrices describing 
the effective spin 1/2. 
For electron systems, these matrices represent the electron spin, while for 
holes which have the internal angular momentum  $J=3/2$, the  matrices describe
the two level heavy hole subsystem $J_z=\pm 3/2$, with light holes, 
$J_z=\pm 1/2$, having a significantly higher energy.

Derivation of the kinematic form of the interaction (\ref{spinorbit1})
is straightforward. In the case of electrons the only possible kinematic 
structure is the Rashba interaction \cite{Bychkov1984},
${\cal H}_{so} \propto ({\bm n}\cdot[{\bm p}\times{\bm S}])$,
where ${\bm S}$ is electron spin and ${\bm n}$ is the unit vector
orthogonal to the plane of 2DEG. This results in Eq.(\ref{spinorbit1})
with $n=1$ and with a real coefficient $\gamma=\gamma_1$.

For holes in an in-plane magnetic field ${\bm B}$ the only possible kinematic 
structure is 
${\cal H}_{so} \propto ({\bm J}\cdot {\bm p})^2({\bm J}\cdot{\bm B})$.
To project this Hamiltonian to the $J_z=\pm 3/2$ heavy hole subspace one 
has to replace $J_+^3 \to \sigma_+$, $J_-^3 \to \sigma_-$. All other powers of
$J$ are projected to zero. Hence Eq.(\ref{spinorbit1}) has
$n=2$ and a complex coefficient $\gamma=\gamma_2 e^{-i\varphi}$,
where $\gamma_2$ is real and proportional to the magnetic field. The phase 
$\varphi$ is the angle of the in-plane magnetic field with respect to the
y-axis, $B_x=-B_{\parallel}\sin\varphi$, $B_y=B_{\parallel}\cos\varphi$,
see Fig. \ref{F1}.

The only possible kinematic structure for the heavy holes Rashba-like 
interaction is  ${\cal H}_{so} \propto ({\bm J}\cdot {\bm p})^2
({\bm n}\cdot[{\bm p}\times{\bm J}])$.
Again, projecting it to the $J_z=\pm 3/2$ subspace with the replacements 
$J_+^3 \to \sigma_+$, $J_-^3 \to \sigma_-$ we come  to Eq.(\ref{spinorbit1})
with $n=3$ and with a real coefficient $\gamma=\gamma_3$,

An odd $n$ in Eq.(\ref{spinorbit1}) implies the change of sign under inversion,
therefore these interactions are due the lack of inversion symmetry in the 
semiconductor heterostructure. In Rashba-like interactions the inversion 
asymmetry is described by the vector ${\bm n}$.
Alternatively, the n=3 case can originate from the Dresselhaus 
SO interaction which is due to the lack of inversion
symmetry in the bulk of a zinc blend semiconductor \cite{Winkler2003}.
In this case the magnitude and the phase of $\gamma$ depends on the
orientation of the heterostructure with respect to crystal axes.

The coefficient $\gamma$ in Eq.(\ref{spinorbit1}) can be presented as
\begin{eqnarray}
&&\gamma = \gamma_ne^{-i\varphi}\nonumber\\
&&\varphi=0 \ \ \  if \ \ \ n=1,3 \nonumber\\
&&\varphi\ne 0 \ \ \ if \ \ \ n=2 \nonumber\\
&&\gamma_n={\tilde \gamma_n}\frac{\epsilon_F}{k_F^n} \ , \ \ \ \ \ \ 
k_F=\sqrt{2 m \epsilon_F}\ ,
\label{gm}
\end{eqnarray}
where $\epsilon_F$ is the Fermi energy (chemical potential).
The nonzero phase for $n=2$ is determined by the orientation of the in-plane 
magnetic field as it is explained above.
The dimensionless coefficient ${\tilde \gamma_n}$ represents
the value of the SO interaction at $p=k_F$ in units of the Fermi energy.
For the Rashba-like interaction (lack of inversion asymmetry in the 
heterostructure) at $n=1,3$ the coefficient can  as large as 
$|{\tilde \gamma_n}| \sim 0.1-0.2$, dependent on the heterostructure 
quantum well.
For the $n=2$, the interaction  induced by an in-plane magnetic field, in GaAS
the value of the coefficient can be a large as
$|{\tilde \gamma_2}| \sim 0.02/Tesla$ \cite{Tommy}, depending on the particulars of the
heterostructure.
The Dresselhaus interaction results is a relatively small value of 
$|{\tilde \gamma_3}|$, in GaAs $|{\tilde \gamma_3}| \sim 0.01$.
It is worth noting that in this case the phase can be nonzero.

\begin{figure}[t!]
     {\includegraphics[width=0.4\textwidth]{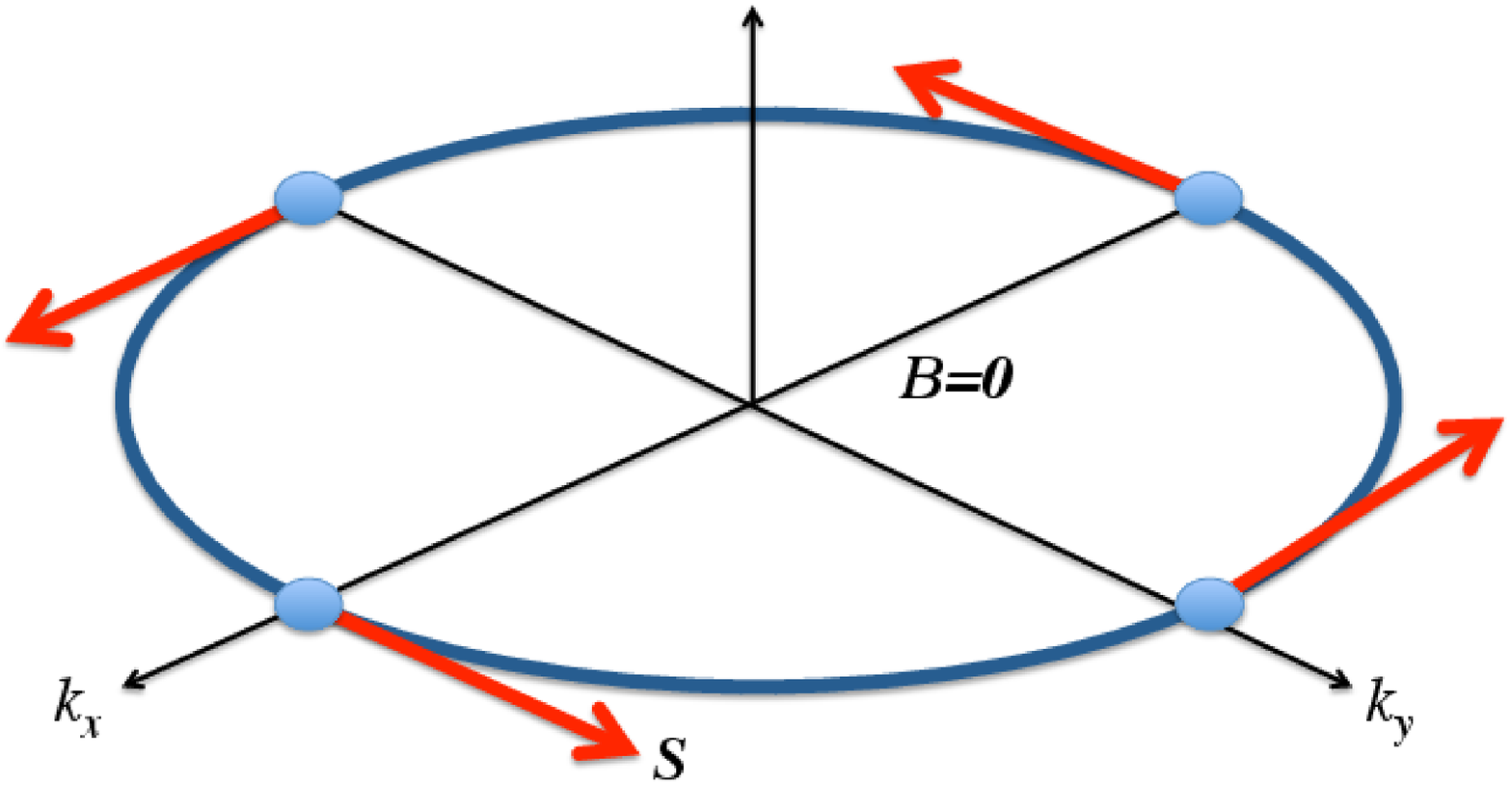}}
     {\includegraphics[width=0.4\textwidth]{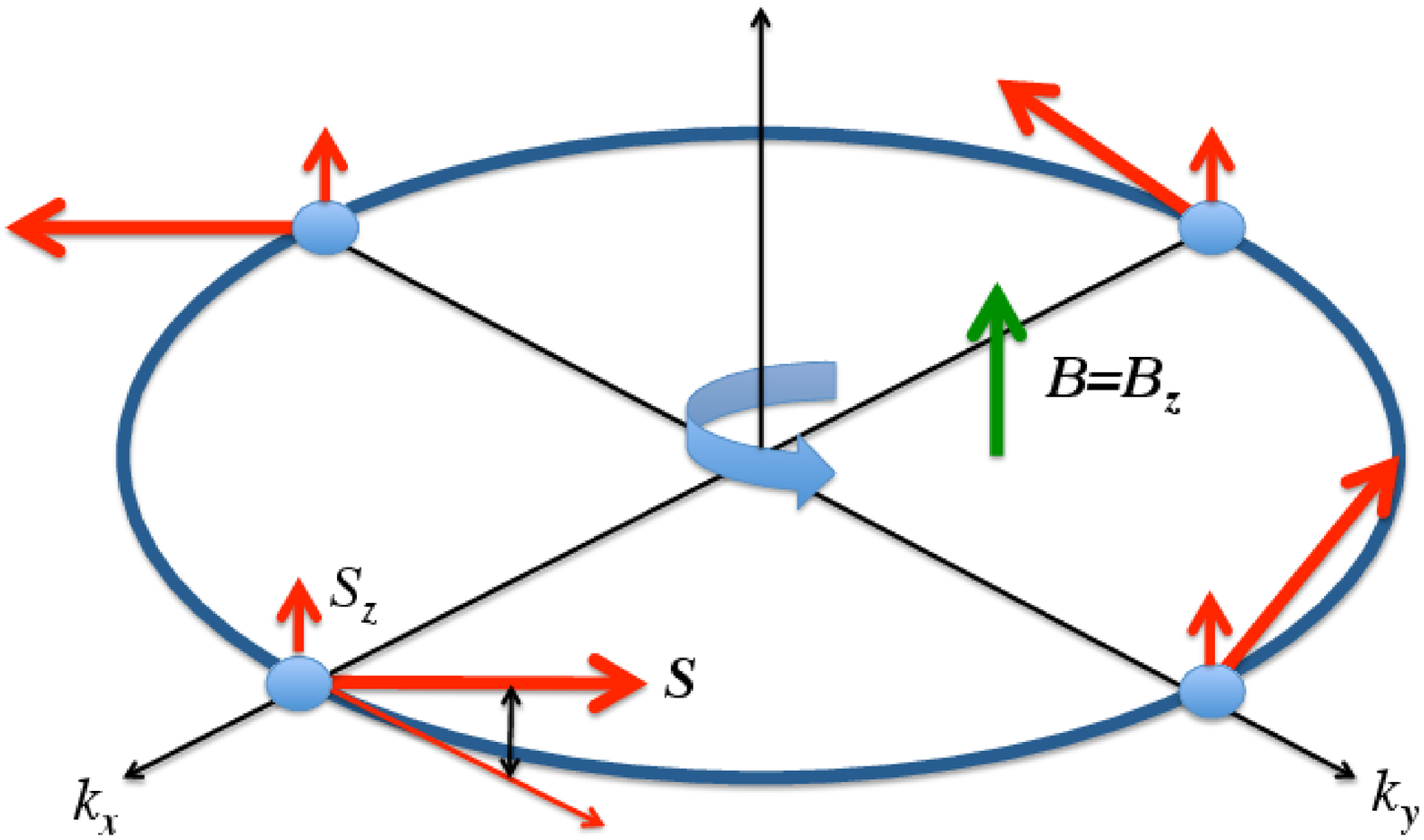}}
	\caption{The spin orientation for linear in momentum
Rashba interaction in the $k$ space.
(a) In the absence of a focusing magnetic field. A particle has a definite 
momentum,
so different points in the k-space correspond to different particles.
(b) A particle in the focusing magnetic field. Both the particle momentum 
and the particle spin evolves in time. Importantly, there is a nonzero
z-component of spin.}
\label{spin orientation}
\end{figure}

We can consider the SO interaction as a momentum dependent effective Zeeman
magnetic field, ${\cal B}({\bf k})$. From (\ref{spinorbit1}),
\begin{eqnarray}
&&{\cal H}_{so} = -{\cal B}\cdot {\bm \sigma}\\
&&{\cal B}=\gamma_n k^n(-\sin (n\theta+\varphi),\cos (n\theta+\varphi),0) 
\nonumber\\
&&{\bf k} = k(\cos \theta, \sin \theta,0) \ ,\nonumber
\label{spinorbit11}
\end{eqnarray}
where $\theta$ is the axial angle in the $k$-space. 
In the absence of a focussing out-of-plane magnetic field electrons 
(holes) move in straight lines, and the spin is parallel (or anti-parallel) 
to the SO effective magnetic field,
\begin{eqnarray}
&&\langle \chi_s| {\bm \sigma}|\chi_s \rangle= 
s(-\sin (n\theta+\varphi),\cos (n\theta+\varphi),0)\\
&&\psi_{{\bf k} s} ({\bf x}) \propto e^{i {\bf k \cdot r}} \chi_s\ , \ \ \ 
\chi_s=\begin{pmatrix}
i e^{i (n \theta+\varphi)} \\
-s
\end{pmatrix}\nonumber
\label{spins}
\end{eqnarray}
where $s=\pm 1$ is the spin polarization, $\psi_{{\bf k} s}$ is the particle
wave function, and $\chi_s$ is the spin wave function.
Note that the effective spin rotates $n$-times when  $k$ rotates only once
in the momentum space. For $n=1$ this is illustrated in 
Fig. \ref{spin orientation}a.
This is why we say that the Hamiltonian (\ref{spinorbit1}) has a given winding number.
A combination of Hamiltionians with different values of $n$ obviously violates this
property.
It is worth noting that for $n=1$ the winding number represents the rotation
of the real spin. However, this is not true for higher values of $n$.
The point is that for $n=2,3$  the effective spin $\sigma$ describes phases
of the heavy hole components $J_z=\pm 3/2$. The expectation value of 
$\langle {\bm \sigma}\rangle$ is not equal to the expectation value of 
${\bm J}$.
For example for $n=2$ $\langle {\bm J}\rangle$ does not change its direction, 
it is directed along ${\bm B}_{\parallel}$, so it does not wind.
Nevertheless, for our purposes the effective spin matters, and therefore the 
classification by the winding numbers makes sense and is important.

The full Hamiltonian, in a transverse magnetic field is
\begin{equation}
{\cal H} = \frac{\pi^2}{2 m} + \left(\frac{i \gamma}{2}\pi^n_- \sigma_+ +h. c. 
\right) -\frac{1}{2} g \mu_B B_z \sigma_z
\label{hamiltonian}
\end{equation}
where ${\bm \pi} = {\bm p} - e {\bm A}$ with  ${\bm A}$ being the vector potential
corresponding to the focusing field ${\bm B}=B_z$, ${\bf B = \nabla \times A}$. 
Besides the vector potential we include Zeeman coupling to the focussing field, 
$g$ is the $g$-factor for the electron (or hole). 
Hereafter we include the Zeeman coupling in the effective SO magnetic field,
\begin{equation}
\label{Bg}
{\cal B} \to \gamma_n k^n(-\sin (n\theta+\varphi),\cos(n\theta+\varphi),0)+\frac{1}{2}
g\mu_B B_z(0,0,1) \ ,
\end{equation}
yielding a finite $z$ component of ${\cal B}$.

When a weak transverse magnetic field is applied, the trajectory of the charge 
carriers is curved, rotating the states through some angls $\theta(t)=\omega t$, after some time $t$, ${\bm k}\to k(\cos\omega t, \sin \omega t, 0)$.
For such a curved trajectory, the SO field is no longer constant, 
but rotates in time, see Eq.(\ref{spinorbit11}).
It is conevenient to perform transformation to the reference frame 
corotating with the SO field
\begin{eqnarray}
\label{tr}
&&\chi_s=U \chi_{s0}\\
&&U(t)= e^{-i\frac{1}{2}\sigma_z (n \theta+\varphi)}, \ \ \ \theta=\omega t \ . \nonumber
\end{eqnarray}
Substitution of (\ref{tr}) in Schroedinger equation
shows that the spin wave function in the corotating frame obeys the following
equation
\begin{equation}
\label{tr1}
i \frac{\partial}{\partial t} U(t) \chi_{s0} = {\cal H} U(t) \chi_{s0}
\end{equation}
\begin{eqnarray}
\label{tr2}
&&i \frac{\partial}{\partial t}\chi_{s0} 
=-{\cal B}_0\cdot{\bm \sigma}\chi_{s0}\\
&&{\cal B}_0\cdot{\bm \sigma}=U^\dagger{\cal B}\cdot 
\boldsymbol{\sigma} U + \frac{1}{2} n \dot{\theta} \sigma_z\nonumber\\
&&{\cal B}_0=(0,\gamma_n k^n,\frac{1}{2} g \mu_B B_z+\frac{1}{2}n\omega) 
\ .\nonumber
\label{berryconnection}
\end{eqnarray}
Solution of  Eq.(\ref{tr2}) is straightforward since ${\cal B}_0$ is time 
independent. There are two eigenstates with energies $\pm |{\cal B}_0|$
corresponding to $s=\pm 1$. In a superposition of these states
the  spin oscillates around  direction of ${\cal B}_0$.
In the eigenstates the effective  spin has the following values
\begin{eqnarray}
\label{sxyz}
&& \langle \sigma_x\rangle=0\nonumber\\
&& \langle \sigma_y\rangle= s\frac{\gamma_n k^n}{|{\cal B}_0|}\nonumber\\
&&\langle \sigma_z\rangle= s\frac{(g \mu_B B_z+n\omega)/2}{|{\cal B}_0|}\ .
\end{eqnarray}
It is worth noting  that the term proportional to $n\omega$ in 
$\langle \sigma_z\rangle$ is a direct consequence of the Corialis force.
Eq. (\ref{sxyz}) gives the effective spin in the corotatine frame. 
In the laboratory frame this gives
\begin{eqnarray}
\label{slab}
&&\left<\sigma_+ \right> = s i e^{i\varphi} \frac{\gamma_n \pi_+^n}{|{\cal B}_0|}\nonumber\\
&&\left<\sigma_z\right> = s\frac{(g \mu_B B_z+n\omega)/2}{|{\cal B}_0|} \ .
\label{spinresults}
\end{eqnarray}
The effective spin evolution in the laboratory frame is illustrated in 
Fig. \ref{spin orientation}b.

Now let us look at the spatial dynamics of the particle in the laboratory frame.
We are interested in a semiclassical wave packet propagating along a
trajectory. To approach the problem we use  the Heisenberg representation, where 
the wavefunctions are time independent, while the operators evolve in time. 
The Heisenberg equations of motion with the Hamiltonian \eqref{hamiltonian}
read 
\begin{eqnarray}
&&\dot{r}_+=v_+ = i[{\cal H}, r_+] = \frac{\pi_+}{m} + 
n i \gamma \pi^{n-1}_- \sigma_+\nonumber\\
&& {\dot{\pi}} _+=i[{\cal H}, \pi_+]=-ieB_zv_+ \ .
\label{velocity2}
\end{eqnarray}
The second of these equations represents the usual Lorentz force, 
interestingly it is not influenced by the SO interaction. On the other hand the
first equations which gives relation between momentum and velocity
depends on the SO interaction.
In the semiclassical limit we have to replace the operators
$\pi_+$, $v_+$, and $\sigma_+$ in Eqs.(\ref{velocity2}) by their expectation 
values. The expectation value of spin is given by Eq.(\ref{spinresults}).
Hence, in the semiclassical limit Eqs.(\ref{velocity2}) are transformed to
\begin{eqnarray}
&&v_{+s} = v_{Fs}e^{i\omega_st}\nonumber\\
&&\pi_{+s} = k_{Fs}e^{i\omega_st}\nonumber\\
&&v_{Fs} =\frac{k_{Fs}}{m}\left(1-\frac{1}{2}s n {\tilde \gamma_n}\right)
\nonumber\\
&&i\omega_sk_{Fs}=im\omega_cv_{Fs} \nonumber\\
&&\omega_c=\frac{-eB_z}{m} > 0 \ .
\label{velocity3}
\end{eqnarray}
Note that while in Eq.(\ref{velocity2}) velocity and momentum are operators,
in Eq.(\ref{velocity3}) these are usual numbers.
Note also that the Fermi momentum and the Fermi velocity depend on spin.
When transforming from (\ref{velocity2}) to (\ref{velocity3}) we 
assume that the SO interaction is small compared to the Fermi energy, 
$\gamma_n \ll 1$, we expand in powers $(s{\tilde \gamma_n})$ keeping only 
the leading term. Note that $(s{\tilde \gamma_n})^2={\tilde \gamma_n}^2$
is independent of spin. Therefore Eq.(\ref{velocity3}) is robust,
the neglected spin dependent term is of the third order,
$\sim s{\tilde \gamma_n}^3$.
On the other hand we assume that the SO interaction is sufficiently strong
to justify adiabatic approximation for spin dynamics, 
${\tilde \gamma_n} \gg \omega_c/\epsilon_F$.
From Eq.(\ref{velocity3}) we immediately find that the frequency of cyclotron
motion depends on spin
\begin{equation}
\label{fs}
\omega_s=\omega_c\left(1-\frac{1}{2}s n {\tilde \gamma_n}\right) \ .
\end{equation}

In a focussing experiment particles are injected at energy equal to the 
chemical potential $\epsilon_F$ independent of spin. However  momentum
of the particle depends on spin. Substitution of the spin expectation
value (\ref{slab}) in the Hamiltonian (\ref{hamiltonian}) gives the
following momentum and velocity
\begin{eqnarray}
\label{vkf}
&&k_{Fs}=k_F\left(1-\frac{1}{2}s {\tilde \gamma_n}\right)\nonumber\\
&&v_{Fs}=\frac{k_F}{m}\left(1+\frac{1}{2}s {\tilde \gamma_n}[1-n]\right) \ .
\end{eqnarray}
Note that both $k_{Fs}$ and $v_{Fs}$ are real, this implies that the momentum
and the velocity are always parallel.
Finally the focusing distance, see Fig.\ref{F1} is
\begin{equation}
\label{L}
L_s=v_{Fs}\frac{2}{\omega_s}=2r_c
\left(1+\frac{1}{2}s {\tilde \gamma_n}\right) \ ,
\end{equation}
where
\begin{eqnarray}
\label{rc}
r_c=\frac{k_F}{m\omega_c}
\end{eqnarray}
is the cyclotron radius.
Remarkably, when expressed in terms of ${\tilde \gamma_n}$
the focussing length is independent of the winding number n.
It does depend on the spin polarisation, and this leads to the double focusing peak. 
For the case of $n=1$, the solution, \eqref{L}, was attained by Z\"ulicke \cite{Zulicke2007}.

The solution  \eqref{velocity3}, \eqref{fs}, \eqref{vkf}
has been obtained in terms of momentum and velocity, see Eq. \eqref{velocity2}.
This solutions does not use explicitly the out-of-plane
component of spin $\langle \sigma_z\rangle$ presented in Eq.\eqref{slab},
the Coriolis force remains ``under the carpet''.
In Appendix \eqref{corialis} we present an alternative solution for $n=1$.
The alternative solution is presented in terms of
the velocity operator and its time derivative. The alternative solution is
essentially based on $\langle \sigma_z\rangle$ and the role of the
Coriolis force is explicit, with the final answer identical to Eqs. \eqref{fs} and \eqref{L}.

\section{Spin-orbit interactions with mixed winding numbers}
\label{sec2}

In this section we consider the general SO interaction.
The Hamiltonian is
\begin{equation}
{\cal H} = \frac{\pi^2}{2 m} -{\cal B}\cdot {\bm \sigma} \ .
\label{hg}
\end{equation}
Again we assume that the spin dynamics are adiabatic, the SO field
is everywhere much larger than the cyclotron frequency, 
$|{\cal B}| \gg \omega_c$. Adiabaticity  is the essential physical 
approximation.
Besides that we assume that  $|{\cal B}| \ll \epsilon_F$.
This assumption is not essential, but it allows us to perform
calculations analytically. Practically the inequality is always valid.
Because of the spin adiabaticity ${\bm \sigma} \to s{\cal B}/|{\cal B}|$
and hence (\ref{hg}) is reduced to the classical Hamiltonian
\begin{equation}
H_{cl} = \frac{\pi^2}{2 m} -s|{\cal B}| \ ,
\label{hcl}
\end{equation}
where ${\cal B}$ is a function of ${\bm \pi}$.
Hence Hamilton Eqs. of motion are
\begin{eqnarray}
\label{heq}
&&v_+=\frac{\partial H_{cl}}{\partial \pi_-}
=\frac{\pi_+}{m}-\frac{s}{|{\cal B}|}\frac{\partial {\cal B}^2}{\partial \pi_-}
\nonumber\\
&&{\dot {\pi}}_+=i\omega_cmv_+ \ ,
\end{eqnarray}
where the z-component in ${\cal B}^2={\cal B}_x^2+{\cal B}_y^2+{\cal B}_z^2$
can be safely neglected.
Let us look for solution in the form
\begin{eqnarray}
\label{kv}
&&\pi_+=ke^{i\theta}\nonumber\\
&&v_+=ve^{i\theta} \ ,
\end{eqnarray}
where $k$ is real. Expanding in powers of $|{\cal B}|/\epsilon_F$ 
Eqs. (\ref{hcl}), (\ref{heq}) are transformed to
\begin{eqnarray}
\label{feq}
&&k=k_F\left(1+s\frac{|{\cal B}|}{2\epsilon_F}\right)\\
&&v=\frac{k_F}{m}\left[1+s\frac{|{\cal B}|}{2\epsilon_F}
-s\frac{k_F}{2|{\cal B}|\epsilon_F}e^{-i\theta}
\frac{\partial {\cal B}^2}{\partial \pi_-}\right]\nonumber\\
&&{\dot \theta}=\omega_c\left[1 
-s\frac{k_F}{4|{\cal B}|\epsilon_F}\left(e^{-i\theta}
\frac{\partial {\cal B}^2}{\partial \pi_-}+
e^{i\theta}
\frac{\partial {\cal B}^2}{\partial \pi_+}\right)\right]\nonumber \ .
\end{eqnarray}
Note that $v$ is generally complex while ${\theta}$ is of course real.
Complex $v$ with real $k$ indicates that the momentum and the velocity are 
generally not parallel.

Integration of Eqs.(\ref{feq}) is straightforward.
To be specific we apply (\ref{feq}) to the physically interesting case
of heavy holes in asymmetric
quantum well and in external in-plane magnetic field of few to several Tesla. 
This experimental setup is frequently employed for the polarisation of 
the QPC. 
The presence of the in-plane magnetic field leads to the simultaneous action of $n=3$ and $n=2$ SO interactions.
\begin{equation}
\label{h32}
{\cal H}_{so} = i \frac{1}{2} \left(\gamma_3 \pi_-^3+\gamma_2 e^{-i\varphi}\pi_-^2
\right) \sigma_+ +h. c.
\end{equation}
In this case
\begin{eqnarray}
\label{bb}
&&{\cal B}^2=\gamma_3^2\pi_-^3\pi_+^3+\gamma_2\gamma_3
\left(e^{i\varphi}\pi_-^3\pi_+^2+e^{-i\varphi}\pi_-^2\pi_+^3\right)+
\gamma_2^2\pi_-^2\pi_+^2\nonumber\\
&&|{\cal B}|=\epsilon_F|{\tilde \gamma_3}| b(\theta)\nonumber\\
&&b(\theta)=\sqrt{1+2({\tilde \gamma_2}/{\tilde \gamma_3})\cos(\theta-\varphi)
+({\tilde \gamma_2}/{\tilde \gamma_3})^2}
\end{eqnarray}
The condition of the spin adiabaticity implies that $b(\theta)$ does not vanish,
$|{\tilde \gamma_3}| b(\theta)\gg \omega_c/\epsilon_F$.
Substitution of (\ref{bb}) in (\ref{feq}) gives the following
\begin{eqnarray}
\label{tv}
&&v_+=\frac{k_F}{m}e^{i\theta}\left[1+s\frac{|{\tilde \gamma_3}|}{2}b(\theta)
-s\frac{3|{\tilde \gamma_3}|}{2}\frac{a(\theta)-ic(\theta)}{b(\theta)}
\right]\nonumber\\
&&{\dot \theta}=\omega_c\left[1-s\frac{3|{\tilde \gamma_3}|}{2}
\frac{a(\theta)}{b(\theta)}\right]\nonumber\\
&&a(\theta)=1+
(5/3)({\tilde \gamma_2}/{\tilde \gamma_3})\cos(\theta-\varphi)
+(2/3)({\tilde \gamma_2}/{\tilde \gamma_3})^2\nonumber\\
&&c(\theta)=({\tilde \gamma_2}/{\tilde \gamma_3})\sin(\theta-\varphi) \ .
\end{eqnarray}
The velocity is not necessarily aligned with the momentum, due to the spin
orbit interaction. 
Integration of these equations results in the following particle trajectory
\begin{eqnarray}
\label{traj1}
&&\theta=\omega_ct+\theta_0-s\frac{3|{\tilde \gamma_3}|}{2}
\int_{\theta_0}^{\theta}
\frac{a(\theta^{\prime})}{b(\theta^{\prime})}d\theta^{\prime}\nonumber\\
&&x+iy=\frac{k_F}{m\omega_c}\left\{i(e^{i\theta_0}-e^{i\theta})\right.\nonumber\\
&&\left.\ \ \ \ \ \ \ \ \ +s\frac{|{\tilde \gamma_3}|}{2}
\int_{\theta_0}^{\theta}e^{i\theta^{\prime}}
\left[b(\theta^{\prime})+3i\frac{c(\theta^{\prime})}{b(\theta^{\prime})}\right]
d\theta^{\prime}\right\}
\end{eqnarray}
Hete $\theta_0$ is the angle which {\it momentum} of the particle
makes with the x-axis at the point of injection, $t=x=y=0$,
$k_x=k\cos\theta_0$, $k_y=k\sin\theta_0$.
We reiterate again that generally the group velocity of the particle is not
parallel to its momentum. In particular, assume  that the particle is
injected along the x-axis, $\theta_0=0$, $k_y=0$.
This does not imply that $v_y=0$, one can easily find from Eq.(\ref{tv})
that $v_y=-s\tilde \gamma_2\frac{3}{2b(0)}\sin\varphi \ne 0$.

In principle Eq.(\ref{traj1}) solves the problem. However, we still need to
specify what is the initial value of the injection angle $\theta_0$.
Intuitively it is zero, $\theta_0=0$. 
In the next section we put this statement on a solid ground by
considering a model of the QCP injector. 
Below in this Section  we set $\theta_0=0$.

\begin{figure}[t!]
     {\includegraphics[width=0.45\textwidth]{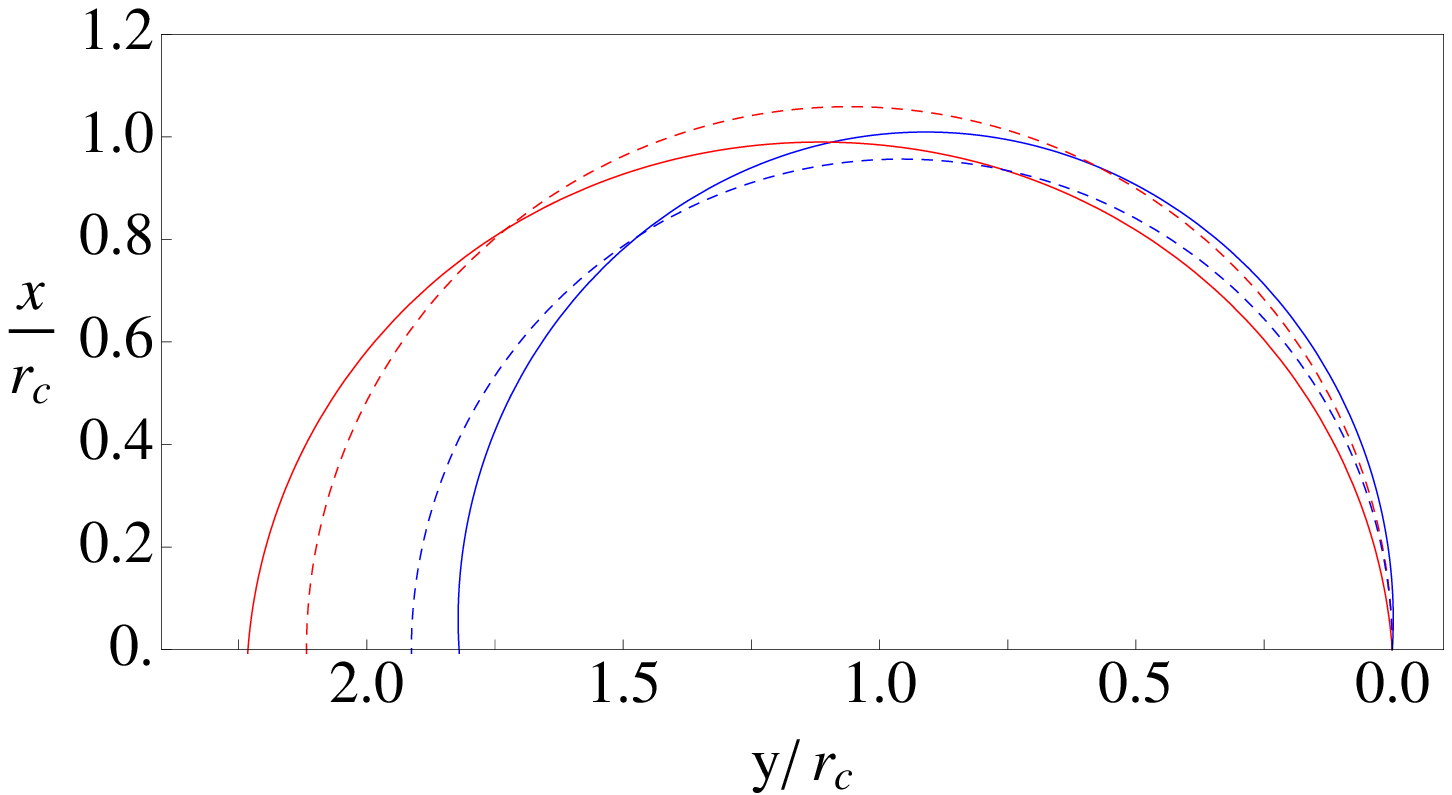}}
     {\includegraphics[width=0.45\textwidth]{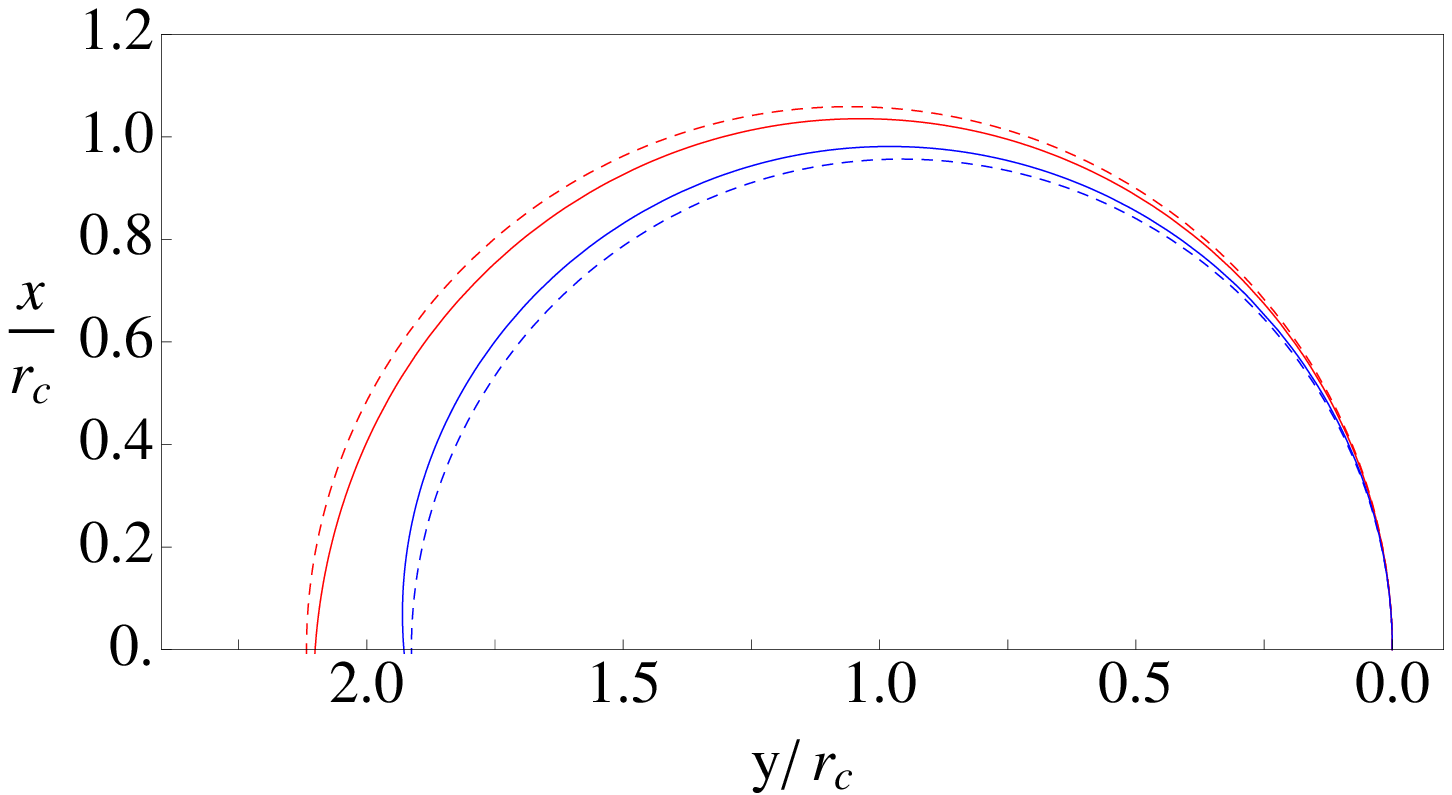}}
     \caption{Particle trajectories in focucing out-of-plane magnetic field.
Red lines correspond to the spin polarization $s=+1$ and blue lines
correspond to the spin polarization $s=-1$.
The value of the ``Rashba parameter'' is ${\tilde \gamma_3}=0.1$.
Coordinates are given in units of cyclotron radius (\ref{rc}), 
the injection point is 
at $x=y=0$.
Dashed lines in both panels are identical, these are
simple semicircles corresponding to zero in-plane magnetic field.
Solid lines account for a non-zero in-plane magnetic field
corresponding to ${\tilde \gamma_2}=0.05$.
Solid lines in the panel a are trajectories for the field parallel
to the direction of injection, $\varphi=-\pi/2$.
Solid lines in the panel b are trajectories for the field perpendicular
to the direction of injection, $\varphi=0,\pi$.
}
\label{trtr}
\end{figure}

Numerical evaluation of Eq.(\ref{traj1}) is straightforward.
Trajectories for ${\tilde \gamma_3}=0.1$, $s=\pm 1$ and
zero in-plane magnetric field, ${\tilde \gamma_2}=0$,
are shown in panel a of Fig.\ref{trtr} by dashed 
blue and red lines.
These are simple semicircles as we discussed in Section II. 
By solid blue and red lines in the same panel a of  Fig.\ref{trtr} we 
show $s=\pm 1$ trajectories for ${\tilde \gamma_3}=0.1$ and 
${\tilde \gamma_2}=0.05$.
A strong dependence of the focusing ditance on the in-plane magnetic field
is evident.
The trajectories depend on direction of the in-plane magnetic field,
Fig.\ref{trtr}a corresponds to $\varphi=-\pi/2$, the field is parallel
to the direction of injection.
In the panel b of Fig.\ref{trtr} we show trajectories for magnetic
field perpendicular to the injection direction, $\varphi=0,\pi$.
The dashed lines are identical to that in Fig.\ref{trtr}a, 
${\tilde \gamma_3}=0.1$ and  ${\tilde \gamma_2}=0$ (zero magnetic field).
The solid lines account for the magnetic field corresponding to
${\tilde \gamma_2}=0.05$. We see that for this orientation of the 
in-plane magnetic
field the focusing distance is independent of the field.

\begin{figure}[t!]
     {\includegraphics[width=0.45\textwidth]{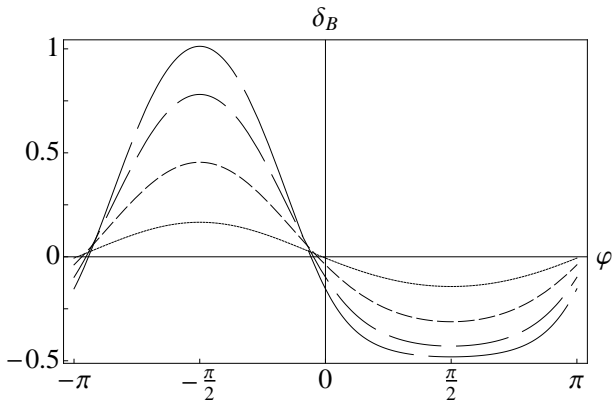}}
     \caption{Dependence of the focusing distance on the
in-plane magnetic field.  $\delta_B$ is the relative field contribution 
to the focusing distance (relative to Rashba).
 The plots show $\delta_B$
versus the field orientation angle $\varphi$
 for the folowing values of the ratio  of the 
field spin-orbit parameter over the Rashba spin-orbit parameter
${\tilde \gamma_2}/{\tilde \gamma_3}=0.1, 0.25, 0.4, 0.5$. 
At sufficiently small $\tilde\gamma_2/\tilde\gamma_3$ the angular
dependence is approximately sinusoidal.
}
\label{trtr1}
\end{figure}

Let us take the focusing shift due to the Rashba interaction, Eq.(\ref{L}),
\begin{equation}
\Delta L_{R}=sr_c{\tilde \gamma_3}
\end{equation}
as a reference value.
Then the total focusing shift is
\begin{equation}
\Delta L=\Delta L_{R}(1+\delta_B) \ ,
\end{equation}
where $\delta_B$ is due to the in-plane magnetic field.
Plots of $\delta_B$ versus the field orientation angle
$\varphi$ for ${\tilde \gamma_2}/{\tilde \gamma_3}=0.1, 0.3, 0.5$
are presented in  Fig.\ref{trtr1}.
One can  check analyticaly that at 
${\tilde \gamma_2} \ll {\tilde \gamma_3}$
the field correction is
$\delta_B\approx -\pi\frac{\tilde \gamma_2}{\tilde \gamma_3}\sin\varphi$.

\section{Injection from QPC}
\label{sec3}

\begin{figure}[t!]
     {\includegraphics[width=0.15\textwidth]{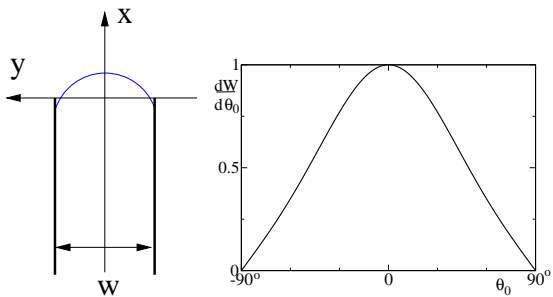}}
     {\includegraphics[width=0.25\textwidth]{Fig3R.eps}}
     \caption{Left panel: A model of the injector as a 1D quantum channel.
Right panel angular distribution of injected electrons/holes.}
\label{qpc}
\end{figure}

The physics of QPCs is incredibly rich, however, for the purposes of this work,
we are interested in the peak momentum at the QPC exit. For this, it is sufficient to 
consider a greatly oversimplified model, excluding all the complexities of QPC physics. 
We consider the injector as a 1D channel aligned with the x-direction.
The width of the channel is $w$ and only one transverse mode is excited in the
channel, see left panel in Fig.\ref{qpc}.
The orbital wave function inside the 1D channel is
\begin{eqnarray}
\label{1dc}
&&\psi \propto e^{ik_xx}cos\varkappa y\ , \ \ \ -w/2< y < w/2\nonumber\\
&&\varkappa=\frac{\pi}{w}
\end{eqnarray}
We assume that $k_x\ll \varkappa$, hence, since the total energy
is equal to Fermi energy, we conclude that $\varkappa \approx k_F$.
The Hamiltonian inside the channel reads
\begin{eqnarray}
\label{h1d}
H&\approx& \frac{k_x^2}{2m}+\frac{\varkappa^2}{2m}\nonumber\\
&-&\epsilon_F
\left[\left(3{\tilde \gamma_3}\frac{k_x}{k_F}
+{\tilde \gamma_2}\cos\varphi\right)\sigma_y
+{\tilde \gamma_2}\sin\varphi\sigma_x\right]
\end{eqnarray}
The spin orbit interaction given by the second line in this
equation determines the spin polarisation inside the channel.

Following the Huygens principle the $k_y$-momentum distribution
after emission from the injector is determined by the Fourier component \cite{Saito1992}
of (\ref{1dc}).
\begin{equation}
\label{dist}
\frac{dW}{dk_y}\propto \left|\int_{-w/2}^{w/2}dy e^{ik_yy}cos\varkappa y\right|^2 \ .
\end{equation}
Converting this to angular distribution we find
\begin{equation}
\label{dist1}
\frac{dW}{d\theta_0}\propto \cos\theta_0
\left|\frac{\cos\left(\frac{\pi}{2}\sin\theta_0\right)}{1-\sin^2\theta_0}
\right|^2 \ .
\end{equation}
The most important conclusion is that the distribution (\ref{dist1}
plotted in the right panel of Fig.\ref{qpc} is peaked at $\theta_0=0$.
This what we used in the previous section.
One can say that this conclusion trivial, but it is not quite.
According to Eq.(\ref{tv}) this implies that the velocity distributions
for injected particles with $s=\pm 1$ are peaked at nonzero angles
\begin{equation}
\label{angle}
\theta_{max}=-\frac{3}{2}s{\tilde \gamma_2}\frac{\sin\varphi}{b(0)} \ .
\end{equation}

The derivation thus far takes no account of adiabatic opening of the QPC, 
reflecting the geometry chosen for this analysis. 
A more realistic geometry includes the adiabatic opening of the QPC, from the
minimum width of the constriction, $W_{min}$, to the point at which adiabatic 
transport breaks down, $W_{max}$.
At the minimum width of the channel, $k_y = \pi/W_{min}$.
As the transport through the QPC is adiabatic, at the exit, the transverse momentum
is $k_{y, x=0} = \pi/W_{max}$, reducing the angular spread of the injected electron/holes.\cite{Molemkamp1990}
The magnitude of this collimation effect depends on the ratio between $W_{min}$ 
and $W_{max}$, and hence the details of the experimental device.
In the above geometry, $W_{min} = W_{max}$ and hence $k_y W_{max} \sim 1$, 
leading to a very large angular spread, with no diffraction fringes. 
We note that while flaring can add significant collimation to the distribution, 
it does not effect peak, which will still be at $k_y=0$, with non-zero velocity, 
as presented in \eqref{angle}.

\section{Summary} 
We consider magnetic focusing of electrons/holes in presence of a
 strong spin orbit interaction. The spin orbit interaction is considered 
in adiabatic approximation, spin follows the effective magnetic field.
We classify spin orbit interactions by winding number $n$, 
the number of rotations
of the effective spin along a close path particle trajectory.
First we consider the case of a singular winding number and extend
previously known result for $n=1$ (usual Rashba interaction) to larger 
values of n, especially to experimentally relevant cases $n=2,3$.
We show that in the case of a singular winding number the particle trajectory
is circular with radius dependent of the spin orbit interaction.
Second we consider the most interesting case of a combination of
spin orbit interactions with different winding numbers.
Particle dynamics in this case are significantly more complex.
We derive general semiclassical equations of motion linearized in
spin orbit interaction. Using the developed technique we consider
in details the case of a combination
of $n=3$ (cubic Rashba) and $n=2$ (in-plane magnetic field).
This analysis can be relevant to dynamics of holes in semiconductors
and provides a general approach to the problem of spin orbit dynamics
while ever the adiabatic approximation is valid. We predict that the application 
of a large in-plane magnetic field will have a significant effect on the magnetic focussing
spectrum, and will be very sensitive to the field orientation.

\section{Acknowledgements}
We thank Tommy Li, Alex Hamilton, Scott Liles, Dmitry Miserev, Alex Milstein, Uli Zuelicke, and Yaroslav Kharkov for important stimulating discussions.

\appendix

\section{Alternative solution of $n=1$ case in terms of velocity operator}
\label{corialis}
Our solution in Section II is obtained in terms of momentum and velocity,
this method is technically the simplest one.
An alternative method is to work with velocity without referring to
momentum. Intuitively this method is more natural in the semicalssical limit.
Here we present the alternative solution for usual, linear in momentum,
Rashba interaction.  Of course final answers of both methods
are identical

We start with usual Rashba spin orbit interaction, 
\begin{eqnarray}
\label{hamil}
{\cal H} = \frac{\boldsymbol{\pi}^2}{2 m } + \gamma \boldsymbol{\pi} \left(\boldsymbol{\sigma} \times \hat{\bf z} \right)-\frac{1}{2} g \mu_B B_z \sigma_z
\end{eqnarray}
which corresponds to the $n=1$ for Eq. \eqref{spinorbit1}. 
The velocity is 
\begin{eqnarray}
\label{vvv}
{\bf v} = \frac{\bf p}{m} + \gamma_1 \boldsymbol{\sigma} \times \hat{\bf z} \ 
\ ,  
\end{eqnarray}
and hence up to terms quadratic in the spin orbit intearction the energy is
\begin{eqnarray}
\label{ee}
{\cal E} = \frac{m {\bf v}^2}{2} -\frac{1}{2} g \mu_B B_z \sigma_z \ .
\end{eqnarray}
Heisenberg equation of motion for velocity reads
\begin{eqnarray}
\label{eqmotionrashba}
{\dot{\bf v}} = \omega_c {\bf v \times \hat{z}} + 
\left\{{\bf v \times \hat{z}}, \boldsymbol{\sigma} \cdot \hat{\bf z}\right\} +  g \mu_B  ({\bf B \cdot \hat{z} }) \gamma_1  \boldsymbol\sigma
\end{eqnarray}
We note the presence of a spin dependent term in the right-hand-side
of the second equation, which has been considered as a 'spin force', 
in analogy with the usual Lorentz, force (the first term) \cite{Shen2005}. 
The third term is the result of the coupling between the spin-orbit component of 
velocity, and the Zeeman interaction with the out of plane field. 
We stress that the 'spin force' appears only within this technique, it
does not appear within the momentum-velocity technique considered in 
the main text, Eqs.(\ref{velocity2}),(\ref{heq}).
In semiclassical approximation the 'spin force' term can be decoupled as
$\left\{ {\bf v \times \hat{z}}, \boldsymbol{\sigma} \cdot \hat{\bf z} \right\}
\to 2 [{\bf v \times \hat{z}}]\langle \sigma_z\rangle$.
Due to the 'spin force' the out-of-plane
spun projection  $\sigma_z$, which was ``under the carpet'' in the main text
solution, here plays an explicit role.
Substitution of $\langle \sigma_z\rangle$ from Eq.(\ref{slab})
gives
\begin{eqnarray}
\dot{\bf v} = \omega_c \left(1 - \frac{s}{2} \tilde\gamma_1 \right) {\bf v \times \hat{z}}
\end{eqnarray}
Hence we get a corrected cyclotron frequency identical to Eq.(\ref{fs})
with $n=1$. 
We note that the additional spin dependent term due to the Zeeman effect
does not result in any finite contribution to $\dot{\bf v}$ in the absence
of the rotation induced correction to $\langle \sigma_z \langle$.
Due to Eq.(\ref{ee}) the particle speed is spin independent, $|{\bf v}|=k_F/m$
and we arrive to the same answer as that in Section II with $n=1$.

\end{document}